\documentclass[12pt]{article} \textheight=24 true cm
\usepackage{graphicx}
\usepackage{amsmath}
\usepackage{amssymb}
\begin{document}
 \begin{center}
{\Large\bf Primordial nucleosynthesis in higher dimensional
cosmology
}\\[8 mm]

 S. Chatterjee\footnote{Relativity and Cosmology Research Centre, Jadavpur University,
Kolkata - 700032, India, and also at IGNOU, New Alipore College,
Kolkata  700053, e-mail : chat\_ sujit1@yahoo.com}\\

\end{center}
\begin{large}
\begin{abstract}

 We investigate nucleosynthesis and element formation in the early
 universe in the framework of higher dimensional cosmology. For
 this purpose we utilize a previous solution of the present
 author, which may be termed as the generalized
 Friedmann-Robertson-Walker model. We find that temperature decays
 less rapidly in higher dimensional cosmology, which we believe
 may have nontrivial consequences \emph{vis-a-vis} primordial
 physics.
\end{abstract}

 ~~~Keywords :  cosmology; higher dimensions;
nucleosynthesis

~~ PACS: 04.20, 04.50 +h

\section{INTRODUCTION}

Big Bang nucleosynthesis begins about three minutes after the Big
Bang, when the universe has cooled down sufficiently to form
stable protons and neutrons, after baryogenesis. The relative
abundances of these particles follow from simple thermodynamical
arguments, combined with the way that the mean temperature of the
universe changes over time (if the reactions needed to reach the
thermodynamically favoured equilibrium values are too slow
compared to the temperature change brought about by the expansion,
abundances will remain at some specific non-equilibrium value).
Combining thermodynamics and the changes brought about by cosmic
expansion, one can calculate the fraction of protons and neutrons
based on the temperature at this point. The answer is that there
are about seven protons for every neutron at the beginning of
nucleogenesis, a ratio that would remain stable even after
nucleogenesis is over. This fraction is in favour of protons
initially primarily because lower mass of the proton favors their
production. Free neutrons also decay to protons with a half-life
of about 15 minutes, and this time-scale is too short to affect
the number of neutrons over the period in which BBN took place,
primarily because most of the free neutrons had already been
absorbed in the first 3 minutes of nucleogenesis-- a time too
short for a significant fraction of them to decay to protons. One
feature of BBN is that the physical laws and constants that govern
the behavior of matter at these energies are very well understood,
and hence BBN lacks some of the speculative uncertainties that
characterize earlier periods in the life of the universe. Another
feature is that the process of nucleosynthesis is determined by
conditions at the start of this phase of the life of the universe,
making what happens before irrelevant. As the universe expands, it
cools. Free neutrons and protons are less stable than helium
nuclei, and the protons and neutrons have a strong tendency to
form helium-4. However, forming helium-4 requires the intermediate
step of forming deuterium. At the time at which nucleosynthesis
occurs, the temperature is high enough for the mean energy per
particle to be greater than the binding energy of deuterium;
therefore any deuterium that is formed is immediately destroyed (a
situation known as the deuterium bottleneck). Hence, the formation
of helium-4 is delayed until the universe becomes cool enough to
form deuterium (at about T = 0.1 MeV), when there is a sudden
burst of element formation. Shortly thereafter, at twenty minutes
after the Big Bang, the universe becomes too cool for any nuclear
fusion to occur. At this point, the elemental abundances are
fixed, and only change as some of the radioactive products of BBN
(such as tritium) decay. Excellent reviews on this topic may be
had in the titles of say, Kolb and Turner ~ \cite{wien} and
 Raychowdhuri \cite{akr}\\
On the other hand the long time goal of unification of gravity
with other forces of nature continues to remain elusive in quantum
field theory. Most recently efforts in this search have been
directed in studying theories where the dimensions of the
spacetime is larger than the (3+1) that we observe
today~\cite{sc}. In the cosmological context the higher
dimensional spacetime is particularly important because Einstein's
field equations generalized to higher dimensions admit solutions
where as the `usual' 3D space expands the extra dimensions shrink
with time such that at a certain stage(may be at the planckian
time)the extra dimensions are no longer visible with the present
day experimental techniques and the cosmology looks effectively
four dimensional. The study of element formation in the framework
of higher dimensional cosmology is particularly relevant in the
sense that both higher dimensional spacetime  and primordial
element formation are important in the early universe.

\section{ The FIELD EQUATIONS}

We have shown earlier~ \cite{mnras} that if we start with a
$(n+2)$-dim spherically symmetric line element as
\begin{equation}
ds^{2} = A^{2}dt^{2}- B^{2}dr^{2}- C^{2}dY_{n}^{2}
\end{equation}
where A, B and C are functions of r and t and
\begin{equation}
dY_{n}^{2}= d\theta_{1}^{2}+ sin^{2}\theta_{1}~ d\theta_{2}^{2}+
sin^{2}\theta_{1}~ sin^{2}\theta_{2}...sin^{2}~\theta_{n-1}
^{2}~d\theta_{n}^{2}
\end{equation}
 and then demand that
the energy momentum tensor should be homogeneous then the above
line element reduces to
\begin{equation}
ds^{2}= dt^{2}-\frac{R(t)^{2}}{(1+ kr^{2}/4)^{2}}(dr^{2}+ r^{2}d
Y_{n}^{2})
\end{equation}
where $k$ is the (n+1) space curvature. This metric form may be
looked as the generalized Friedmann-Robertson-Walker universe.We
studied this metric form earlier to get an interesting
astrophysical observation that the mean density of any local
inhomogeneity in this generalized FRW universe must be equal to
the mean cosmological density. In this report we turn our
attention to the vexed problem of nucleosynthesis in the early
universe.\\
As pointed out earlier higher dimensional spacetime is
particularly relevant to the early universe and so the question of
element formation in the higher dimensional universe is
particularly important. From Einstein's field equations
\begin{equation}
R_{ij} - \frac{1}{2}Rg_{ij}= -T_{ij}
\end{equation}
where $T_{ij}$ stand for the energy momentum tensor appropriate to
our matter field we get for the metric(3) the following field
equations
\begin{equation} \frac{n(n+1)}{2}\frac{\dot{R}^{2}+ k}{R^{2}}=
\rho
\end{equation}
\begin{equation}
-\frac{n\ddot{R}}{R}-\frac{n(n-1)}{2}\frac{\dot{R}^{2}+ k}{R^{2}}=
p
\end{equation}
where $\rho$ and p are the homogeneous mass density and pressure
respectively.In the early universe one takes the radiation
dominated case as our equation of state such that for a (n+2) dim.
universe, $ p = \frac{\rho}{n-1}$. One of the phenomenal successes
favoring bigbang cosmology is the almost correct prediction of the
of the primeval nucleosynthesis, particularly the observed
abundances of the light nuclei in the current universe.\\ Now it
can be shown via the field equations(3) and (4) that for the
radiation dominated case
\begin{equation}
\rho = \frac{2n(n+1)}{(n+2)^{2}}(\frac{1}{t^{^{2}}})
\end{equation}
Assuming absence of any dissipative mechanisms(for example,
viscosity, friction etc.) and also that the laws of thermodynamics
are valid in the early universe also with such huge temperature
one gets from elementary thermodynamical considerations ~
\cite{alvarez} that
\begin{equation}
\rho = \sigma T_{rad}^{n+2}
\end{equation}
where $\sigma$  is the higher dimensional Steffan's constant,
whence it follows that \begin{equation} T_{rad}=
\frac{2n(n+1}{(n+2)^{2}\sigma}t^{-\frac{2}{n+2}}
\end{equation}
where $T_{rad}$ is the temperature of radiation and `t' is the age
of the universe. For the usual 4D spacetime it reduces to the
wellknown relation \\
$T_{kelvin}= 1.52\times 10^{10} t^{-1/2}$ s\\ The equation (8) is
the key equation for our attempt to investigate the effect of
extra dimensions on the process of nucleosynthesis in early
universe. We shall however follow the latter here. We here study
the situation when elementary particles have already materialized
allowing us to take the low temperature approximation, $ T <<
m_{\mu}c^{2}= T_{\mu}$ for their distribution function. Here
$m_{\mu}$ is the mass of a particular species. We here try to
study the equilibrium condition for neutrinos with other species.
As the reactions involving the neutrinos fall within the category
of weak interactions and $T < T_{\mu}$ the cross section of a
typical reaction is of the order of
\begin{equation}
A = f^{2}h^{-4}(kT)
\end{equation}

where f is the weak coupling constant. For simplicity it is
further assumed that the constant,A does not depend on the number
of spatial dimensions.\\
Moreover the number densities of the participating particles(say,
muons)are, for (n+1) spatial dimensions, of the order of
$(T/ch)^{n+1}$ and for reactions involving muons at low
temperatures an exponential damping factor of $exp(-T/T_{\mu})$
should also be considered.\\ In the cosmological context one
should also consider the rate of expansion of the background,
which from equations (4) and (5) give \begin{equation}
H^{2}=\frac{\dot{R}^{2}}{R^{2}}\sim t^{-2} = T^{n+2}
\end{equation}
Thus the ratio of the reaction rate to the expansion rate now
becomes
\begin{equation}
\frac{Q}{H}\sim (
\frac{T}{10^{10}K})^{\frac{n+4}{2}}exp(\frac{10^{12}K}{T})
\end{equation}
One recovers the familiar 4D form when $n=2$, As the temperature
falls below the critical level of$10^{10}K$, the exponential
decays rapidly.One can at this stage call attention to a
significant quantitative difference from the 4D case.From
equations (9) and (12) it is tempting to suggest that as the
temperature falls less rapidly in higher dimensional cosmology
than the analogous 4D situation it takes relatively more time for
the elementary particles to cool below the threshold temperature.
More importantly the quotient $\frac{Q}{H}$ is more sensitive to
temperature fluctuations in multidimensional universe. Thus Q is
larger than H for $T > 10^{12}$ depending on the number of extra
dimensions and in these temperature ranges the neutrinos would be
in thermal equilibrium with the rest of the other species. As the
temperature falls further below $10^{10}K$ both the terms in the
rhs of equation(12) drop rapidly, which means that the reactions
involving neutrinos run at a slower rate than compared to the
expansion of the universe. This triggers the so called decoupling
of the neutrinos from the rest of the other constituents of matter
and as pointed earlier in the higher dimensional universe it takes
relatively more time for this decoupling phase of the neutrinos to
occur.However, the theoretical and observational consequences of
this supposed time lag for the initiation of the decoupling era
need to be worked out in more detail before any definite
inferences could be drawn.\\\\
\textbf{Acknowledgment : }
 The financial support of UGC, New Delhi in the
form of a MRP award is  acknowledged .



\bibliographystyle{aipproc}   


\IfFileExists{\jobname.bbl}{}
 {\typeout{}
  \typeout{******************************************}
  \typeout{** Please run "bibtex \jobname" to optain}
  \typeout{** the bibliography and then re-run LaTeX}
  \typeout{** twice to fix the references!}
  \typeout{******************************************}
  \typeout{}
 }



\end{large}
\end{document}